\documentclass[twocolumn,showpacs,prb,amsmath,amssymb]{revtex4}
\usepackage{graphicx}
\usepackage{dcolumn}
\usepackage{bm}
\usepackage{epsfig}
\usepackage{amsmath}
\usepackage[dvips]{color}
\newcommand{\beann}{\begin{eqnarray*}} \newcommand{\eeann}{\end{eqnarray*}}
\newcommand{\bea}{\begin{eqnarray}} \newcommand{\eea}{\end{eqnarray}}

\newcommand{\DSJ}{DSJ}
\newcommand{\SSJ}{SSJ}
\begin{document}
\title{Molecular junctions in the Coulomb blockade regime: rectification and nesting}
\author{Bo Song}
\author{Dmitry A.~Ryndyk}
\author{Gianaurelio Cuniberti}
\address{Institute for Theoretical Physics, University of Regensburg, D-93040 Germany}
\date{November 7, 2006}
\begin{abstract}
Quantum transport through single molecules is very sensitive to
the strength of the molecule-electrode contact. Here, we
investigate the behavior of a model molecular junction weakly
coupled to external electrodes in the case where charging effects
do play an important role (Coulomb blockade regime). As a minimal
model, we consider a molecular junction with two spatially
separated donor and acceptor sites. Depending on their mutual
coupling to the electrodes, the resulting transport observables
show well defined features such as rectification effects in the
$I$-$V$ characteristics and nesting of the stability diagrams. To
be able to accomplish these results, we have developed a theory
which allows to explore the charging regime via the nonequilibrium
Green function formalism parallel to the widely used master
equation technique. Our results, beyond their experimental
relevance, offer a transparent framework for the systematic and
modular inclusion of a richer physical phenomenology.
\end{abstract}
 \maketitle
\section{Introduction}
\begin{figure}[b]
\centerline
{\psfig{file=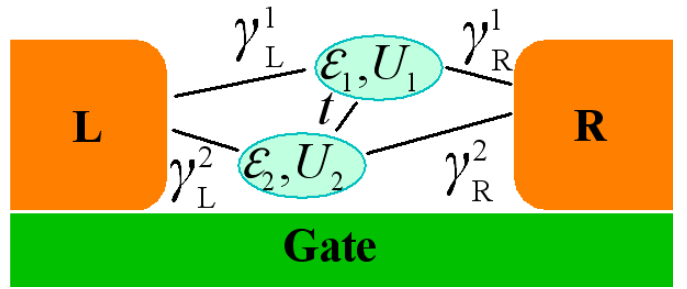,width=1.02\columnwidth}}
\caption{\label{gnrl-DQD}(color) The general configuration of a
double site junction. The levels $\epsilon_{1,2}$ with charging
energies $U_{1,2}$ are connected via $t$ and coupled to the
electrodes via the linewidth injection rates
$\gamma^{i}_{\alpha}$.}
\end{figure}

Single molecule electronics\cite{NR03,JGA00,cfr05} has been mostly
investigated in the high temperature and strong contact to the
electrode regime. The opposite limit of low temperature and weakly
coupled molecular junctions pose a challenge to the currently
available experimental techniques. Still the possibility to probe
the spectroscopy of single molecule junctions via a lateral gate
could offer new insights in the peculiar coupling of the
electrical and mechanical degrees of freedom at the nanoscale. In
order to be able to establish the transport mechanisms governing
such molecular junctions in the Coulomb blockade (CB) regime, a
technique which could tackle on one hand single electron charging
effects and, on the other hand, the inclusion of the
electron-vibron coupling is of extreme importance. The
nonequilibrium Green function (NEGF) formalism has been recently
employed to describe transport observables on the base of a
density functional theory description of the electronic
structure\cite{cfr05,transiesta02,TaylorBS03,gDFTB00,gDFTB02,
gDFTB02-2,smeago06,abinitioNEGF05} and model Hamiltonian
approaches.\cite{GalperinN03,GNR06PRL,Pals96jpcm,datta06condmat}
The NEGF was applied to describe the influence of the vibron
dynamics onto a molecular transistor in the strong coupling
regime\cite{Ryndyk05prb,Ryndyk06prb} but it is typically
substituted with master equation approaches when coming to the
case of weak coupling to the electrodes. Our purpose is to study
the problem of a two site donor/acceptor molecule in the CB regime
within the NEGF as a first step to deal with the phenomenology of
a rigid multilevel island. The nuclear dynamics (vibrations)
always present in molecular junctions could be then modularly
included in this theory. Our method developed in this paper can be
calibrated on the well-studied double quantum dot problem
\cite{DS-review} and could be possibly integrated in density
functional theory based approaches to molecular conductance.
\\
Here, we apply our theory to the case of a two site energetically
asymmetric molecular junction. In the case of serial coupling to
the electrodes, this configuration consists, de facto, in a
molecular rectifier (diode) as proposed long time ago by Aviram
and Ratner\cite{AR74a} and recently experimentally
realized.\cite{Elbing05pnas} We show that the sequential tunneling
regime, being a fundamental different regime from coherent
transport, is compatible to the observed rectification
features.\cite{Elbing05pnas} The serial arrangement of a double
site correlated molecule between two leads is possibly the
simplest configuration. The most general case (see
Fig.~\ref{gnrl-DQD}), which includes parallel pathways, shows in
the sequential tunneling regime an interplay of correlated effect
and interference eventually bringing to the phenomenon of a
nesting of the stability diagrams due to possible different
charging energies.
\\
In this paper, we introduce a powerful \emph{Ansatz} for the NEGF
which is related both to the equation-of-motion (EOM) method and
to the Dyson equation approach. From the knowledge of the Green
function (GF) we then calculate the transport observables. Our
results are of a particular interest in its own at a formal level.
In the case of a single site junction (\SSJ) with Coulomb
interaction (Anderson impurity model), the \emph{conductance}
properties have been successfully studied by means of the EOM
approach in the cases related to CB\cite{MWL91} and the Kondo
effect.\cite{MWL93} Later the same method was applied to some
two-site models.\cite{Niu95prb,Pals96jpcm,Lamba00prb} Multi-level
systems were started to be considered only
recently.\cite{Palacios97prb,Yi02prb} Besides, there are some
difficulties in building the lesser GF in the nonequilibrium case
(at finite bias voltages) by means of the EOM
method.\cite{NiuEOM99jpcm,Swirkowicz03prb,Bulka03prb} Here, we
develop a self-consistent nonequilibrium method for the GF of a
single-site junction (SSJ) and of a double-site junctions (DSJ).
The results of the EOM method could be calibrated with other
available calculations, such as the master equation approach and
the non-crossing approximation.
\\
This paper is organized as follows: after the
derivation of the SSJ and the \DSJ\
nonequilibrium results for the retarded and
lesser GFs (Sec. II), we do show their effects
on the transport observables (Sec. III).
\section{System and method}
The goal of this paper is the determination of the transport
observables for a minimal model of a molecular-junction in the CB
regime, namely a double site correlated impurity Hamiltonian
coupled to extended electrode states. For clarity, we first
describe our method in the more familiar problem of a single site
junction, which is the well-known Anderson impurity model.
\subsection{Single site case}
The Anderson impurity model is used to describe the Coulomb
interaction on a single site: \beann
H=H_{\textrm{D}}+\sum_{\alpha}(H_{\alpha}+H_{\alpha \textrm{D}}),
\eeann where \beann
&&H^{\phantom{\dagger}}_{\textrm{D}}=\sum_{\sigma}\epsilon^{\phantom{\dagger}}_{\sigma}
d^{\dagger}_{\sigma}d^{\phantom{\dagger}}_{\sigma}
+\frac{1}{2}U n_{\sigma}n_{\bar{\sigma}},\\
&&H^{\phantom{\dagger}}_{\alpha}=\sum_{k,\sigma}\epsilon^{\alpha}_{k,\sigma}
c^{\dagger}_{\alpha,k,\sigma}c^{\phantom{\dagger}}_{\alpha,k,\sigma},\\
&&H^{\phantom{\dagger}}_{\alpha \textrm{D}
}=\sum_{k,\sigma}\left(V^{\phantom{\dagger}}_{\alpha,k,\sigma}
c^{\dagger}_{\alpha,k,\sigma}d^{\phantom{\dagger}}_{\sigma}
+V^{*\phantom{\dagger}}_{\alpha,k,\sigma}d^{\dagger}_{\sigma}c^{\phantom{\dagger}}_{\alpha,k,\sigma}\right),
\eeann
where $d$ and $c$ are the operators for
electrons on the dot and on the left
($\alpha=\textrm{L}$) and the right
($\alpha=\textrm{R}$) lead, $U$ is the Coulomb
interaction parameter, $\epsilon_{\sigma}$ is
the $\sigma$ level of the quantum dot, while
$\epsilon^{\alpha}_{k,\sigma}$ are $\sigma$
level of lead $\alpha$ in $k$ space,
$\sigma=\uparrow,\downarrow$.
%
%
With the help of the EOM and a truncation approximation, we can
get a closed set of equations for the retarded and advanced GFs
$G^{\textrm{r}/\textrm{a}}_{\sigma,\tau}$,\cite{HaugJ096}
\begin{subequations}
\begin{align}
\label{eq-GF-SS-1}
&(\omega-\epsilon_{\sigma}-\Sigma^{\textrm{r}/\textrm{a}}_{\sigma})
G^{\textrm{r}/\textrm{a}}_{\sigma,\tau}
=\delta_{\sigma,\tau}+UG^{(2)\textrm{r}/\textrm{a}}_{\sigma,\tau},\\
\label{eq-GF-SS-2}
&(\omega-\epsilon_{\sigma}-U-\Sigma^{\textrm{r}/\textrm{a}}_{\sigma})
G^{(2)\textrm{r}/\textrm{a}}_{\sigma,\tau} =\langle
n_{\bar{\sigma}}\rangle \delta_{\sigma,\tau},
\end{align}
\end{subequations}
where $G^{\textrm{r}/\textrm{a}}_{\sigma,\tau}=\langle\langle
d_{\sigma}
|d^{\dagger}_{\tau}\rangle\rangle^{\textrm{r}/\textrm{a}}$,
$G^{(2)\textrm{r}/\textrm{a}}_{\sigma,\tau}=\langle\langle
n_{\bar{\sigma}}d_{\sigma}|d^{\dagger}_{\tau}\rangle\rangle^{\textrm{r}/\textrm{a}}$
and \bea\label{eq-SS-self-energy}
\Sigma^{\textrm{r}/\textrm{a}}_{\sigma}(\omega)
=\Sigma^{\textrm{r}/\textrm{a}}_{\textrm{L},\sigma}
+\Sigma^{\textrm{r}/\textrm{a}}_{\textrm{R},\sigma}
=\sum_{\alpha,k}{\frac{|V_{\alpha,k,\sigma}|^{2}}
{\omega-\epsilon^{\alpha}_{k,\sigma}\pm \textrm{i}0^{+}}} \eea are
the electron self-energies.
\begin{figure}[t]
\rightline {\psfig{file=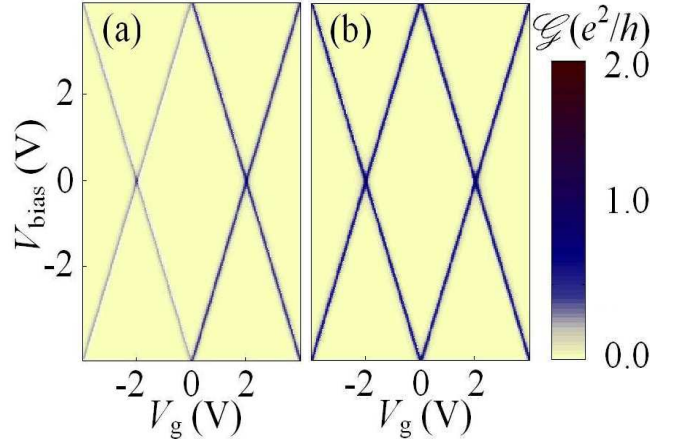, height=6.0cm,
width=1.07\columnwidth} } \caption{\label{stab-diagram-DS}(color)
The stability diagram of a SSJ with $\epsilon_{\sigma}=2.0~$eV,
\mbox{$U=4.0~$eV}, \mbox{$\Gamma_{L}=\Gamma_{R}=0.05~$eV}. (a) The
uncorrect result obtained by means of the widely used formula in
Eq.~(\ref{widely-used-LGF}) for the lesser GF is not symmetric for
levels $\epsilon_{\sigma}$ and $\epsilon_{\sigma}+U$. (b) Results
obtained by means of our \textit{Ansatz} in
Eq.~(\ref{eq:Gkeldysh-SS-new}) shows correctly symmetric for
levels $\epsilon_{\sigma}$ and $\epsilon_{\sigma}+U$.}
\end{figure}
\subsubsection{Mapping on retarded Green functions: equilibrium case}

There are two typical ways to calculate GFs. The first is by means
of the Dyson equation and Feynman diagrams, the second is by means
of the EOM.\cite{Mahan90} For retarded GFs, from the EOM method,
and with the help of Eqs.~(\ref{eq-GF-SS-1}) and
(\ref{eq-GF-SS-2}), we can get \beann
G^{\textrm{r}}&=&G^{\textrm{r}}_{0}+G^{\textrm{r}}_{0}UG^{(2)\textrm{r}}\\
&=&G^{\textrm{r}}_{0}+G^{\textrm{r}}_{0}\Sigma^{\textrm{EOM}}G^{(1)\textrm{r}},
\eeann where $G^{\textrm{r}}$ is single-particle GF matrix \beann
G^{\textrm{r}}=\begin{pmatrix}
G^{\textrm{r}}_{\uparrow,\uparrow} & G^{\textrm{r}}_{\uparrow,\downarrow} \\
G^{\textrm{r}}_{\downarrow,\uparrow} &
G^{\textrm{r}}_{\downarrow,\downarrow}
\end{pmatrix},
\eeann and
$G^{(1)\textrm{r}}_{\sigma,\tau}=G^{(2)\textrm{r}}_{\sigma,\tau}/\langle
n_{\bar{\sigma}}\rangle$. $G^{\textrm{r}}_{0}$ describes the
single-particle spectrum without Coulomb interaction, but
including the effects from the electrodes.
$\Sigma^{\textrm{EOM}}_{\sigma,\tau}=U\langle
n_{\bar{\sigma}}\rangle$ is the Hartree-like self-energy of our
model. Since there is only Coulomb interaction on the site with
the levels $\epsilon_{\sigma}$, the Fock-like self-energy is
vanishing.
\\
Alternatively, by means of the Dyson equation and the second-order
truncation approximation, taking Hartree-like self-energies
$\Sigma^{\textrm{H}}_{\sigma,\tau}=U\langle
n_{\bar{\sigma}}\rangle~(=\Sigma^{EOM}_{\sigma,\tau})$, we can
also get the retarded GFs as follows,\cite{Mahan90} \bea
\label{eq-Dyson-SS-1}
G^{\textrm{r}}=G^{\textrm{r}}_{0}+G^{\textrm{r}}_{0}\Sigma^{\textrm{H}}G^{\textrm{r}}_{1},
\eea where
$G^{\textrm{r}}_{1}=G^{\textrm{r}}_{0}+G^{\textrm{r}}_{0}\Sigma^{\textrm{H}}G^{\textrm{r}}_{0}$
is the first-order truncation GF.
\\
Within the level of the second-order truncation approximation, we
see that there is a map between the EOM results and the Dyson
results:
\begin{subequations}
\label{eq-mapping-SS}
\begin{align}
\label{eq-a}
&&G^{\textrm{r}}~=~G^{\textrm{r}}_{0}~+~G^{\textrm{r}}_{0}~
\Sigma^{\textrm{H}}~G^{(1)\textrm{r}}\hspace{1cm}\textrm{EOM},\\
&&\updownarrow\hspace{3.1cm}\updownarrow\hspace{2.3cm}\nonumber\\
\label{eq-b} &&G^{\textrm{r}}~=~G^{\textrm{r}}_{0}~+~
G^{\textrm{r}}_{0}~\Sigma^{\textrm{H}}~G^{\textrm{r}}_{1}~\hspace{1cm}\textrm{Dyson}.
\end{align}
\end{subequations}
Eqs.~(\ref{eq-mapping-SS}) prompts a way to include further
many-particle effects into the Dyson equation, (Eq.~(\ref{eq-b}),
by replacing the \textit{Dyson-first-order} retarded Green
function $G^{\textrm{r}}_{1}$ with the \textit{EOM} $G^{(1)r}$.
Then one obtains already the correct results to describe CB while
keeping the Hartree-like self-energy.

\subsubsection{Mapping on contour Green functions: nonequilibrium case}

Introducing now the contour GF $\check{G}$, we can get the Dyson
equation as
follows,\cite{Kadanoff62book,Keldysh65,Rammer86RMP,HaugJ096} \bea
\label{eq-Dyson-SS-2}
\check{G}=\check{G}_{0}+\check{G}_{0}\check{\Sigma}\check{G}, \eea
where $\check{\Sigma}$ is the self-energy matrix.\cite{HaugJ096}
\\
According to the approximation for the retarded GF in
Eq.~(\ref{eq-Dyson-SS-1}), we take the second-order truncation on
Eq.~(\ref{eq-Dyson-SS-2}), and then get \beann
\check{G}=\check{G}_{0}+\check{G}_{0}\check{\Sigma}^{\textrm{H}}\check{G}_{1},
\eeann where
$\check{G}_{1}=\check{G}_{0}+\check{G}_{0}\check{\Sigma}^{\textrm{H}}\check{G}_{0}$
is the first-order contour GF, and $\check{G}_{0}$ has already
included the lead effects.
\\
Similar to the mapping in Eq.~(\ref{eq-mapping-SS}), we perform an
\textit{Ansatz} consisting in substituting the
\textit{Dyson-first-order} $G^{\textrm{r}/\textrm{a}/<}_{1}$ with
the \textit{EOM} one $G^{(1)\textrm{r}/\textrm{a}/<}$ to consider
more many-particle correlations, while the \textit{EOM}
self-energy is used for the \textit{Dyson} equation for
consistency: \bea \label{eq:mapping}
\begin{matrix}
\check{G} & = & \check{G}_{0} & + &
\check{G}_{0}&\check{\Sigma}^{\textrm{H}}&\check{G}_{1}
& & & \textrm{Dyson},\\
\updownarrow& & & & & &\uparrow & & &\\
\check{G}& &&&&&\check{G}^{(1)} & & & \textrm{EOM}.
\end{matrix}
\eea Then, using the Langreth theorem,\cite{HaugJ096} we get the
lesser GF, \bea \label{eq:Gkeldysh-SS-new}
G^{<}&=&G^{<}_{0}+G^{\textrm{r}}_{0}\Sigma^{\textrm{H},\textrm{r}}G^{(1)<}
+G^{<}_{0}\Sigma^{\textrm{H},\textrm{a}}G^{(1)\textrm{a}}\nonumber\\
&=&G^{<}_{0}+G^{\textrm{r}}_{0}UG^{(2)<}
+G^{<}_{0}UG^{(2)\textrm{a}} \eea where
$G^{\textrm{r}/\textrm{a}/<}_{0}$ are GFs for $U=0$, \textit{but}
including the lead effects, \textit{i.e.} \beann
&&G^{<}_{0}=g^{<}_{0}+g^{\textrm{r}}_{0}\Sigma^{<}G^{\textrm{a}}_{0}
+g^{<}_{0}\Sigma^{\textrm{a}}G^{\textrm{a}}_{0}
+g^{\textrm{r}}_{0}\Sigma^{\textrm{r}}G^{<}_{0},\\
&&G^{\textrm{r}/\textrm{a}}_{0}=g^{\textrm{r}/\textrm{a}}_{0}
+g^{\textrm{r}/\textrm{a}}_{0}\Sigma^{\textrm{r}/\textrm{a}}G^{\textrm{r}/\textrm{a}}
, \eeann with $g^{\textrm{r}/\textrm{a}/<}_{0}$ the free electron
GF, and \beann \Sigma^{\textrm{r}/\textrm{a}/<}=
\begin{pmatrix}
\Sigma^{\textrm{r}/\textrm{a}/<}_{\uparrow} & 0 \\
0 & \Sigma^{\textrm{r}/\textrm{a}/<}_{\downarrow}
\end{pmatrix},
\eeann
$\Sigma^{<}_{\sigma}=\textrm{i}\sum_{\alpha}{\Gamma_{\alpha}
f_{\alpha}(\omega)}$, and
$\Gamma_{\alpha}=\textrm{i}(\Sigma^{\textrm{r}}_{\alpha}-\Sigma^{\textrm{a}}_{\alpha})$,
$f_{\alpha}(\omega)=f(\omega-\mu_{\alpha})$, $f$ is the
equilibrium Fermi function and $\mu_{\alpha}$ is the
electro-chemical potential in lead $\alpha$;
$\Sigma^{\textrm{r}/\textrm{a}}_{\alpha}$ are the
retarded/advanced electron self-energies from
Eq.~(\ref{eq-SS-self-energy}) and
$G^{(1)\textrm{r}/\textrm{a}/<}_{\sigma,\tau}=G^{(2)\textrm{r}/\textrm{a}/<}_{\sigma,\tau}
/\langle n_{\bar{\sigma}}\rangle$. Performing the same
\textit{Ansatz} on the double-particle GF, from
Eq.~(\ref{eq-GF-SS-2}) we can get \bea \label{GF2-SS}
G^{(2)<}=G^{(2)\textrm{r}}\Sigma^{(2)<}G^{(2)\textrm{a}}, \eea
with $\Sigma^{(2)<}_{\sigma}=\Sigma^{<}_{\sigma}/\langle
n_{\bar{\sigma}}\rangle$.
\\
The lesser GFs in Eq.~(\ref{eq:Gkeldysh-SS-new}) can also be
obtained directly from the general formula\cite{HaugJ096} \beann
\tilde{G}^{<}(\omega)=\tilde{G}^{<}_{0}+\tilde{G}^{\textrm{r}}_{0}
\tilde{\Sigma}^{\textrm{r}}\tilde{G}^{<}
+\tilde{G}^{\textrm{r}}_{0}\tilde{\Sigma}^{<}\tilde{G}^{\textrm{a}}
+\tilde{G}^{<}_{0}\tilde{\Sigma}^{\textrm{a}}\tilde{G}^{\textrm{a}},
\eeann with the help of the \textit{Ansatz} in
Eq.~(\ref{eq:mapping}). It should be noted that
Eq.~(\ref{eq:Gkeldysh-SS-new}) is very different from the lesser
GF formula, \bea\label{widely-used-LGF}
G^{<}=G^{\textrm{r}}\Sigma^{<}G^{\textrm{a}},\eea which is widely
used for both first-principle \cite{transiesta02,smeago06,PdC04}
and model Hamiltonian calculations.\cite{Pals96jpcm} It should be
noted that the self-energy $\Sigma^{<}$ in
Eq.~(\ref{widely-used-LGF}) contains \textit{only} contributions
from the electrodes.
\\
The numerical calculation results of conductance dependence on the
bias and gate voltages by the two different NEGF
Eqs.~(\ref{eq:Gkeldysh-SS-new}) and (\ref{widely-used-LGF}) are
shown in Fig.~\ref{stab-diagram-DS}. As we can see in the left
panel, the adoption of Eq.~(\ref{widely-used-LGF}) results in
incorrectly symmetry-breaking in the gate potential. This wrong
behavior is corrected in the right panel where
Eq.~(\ref{eq:Gkeldysh-SS-new}) has been used.
\begin{figure}
\centerline{\psfig{file=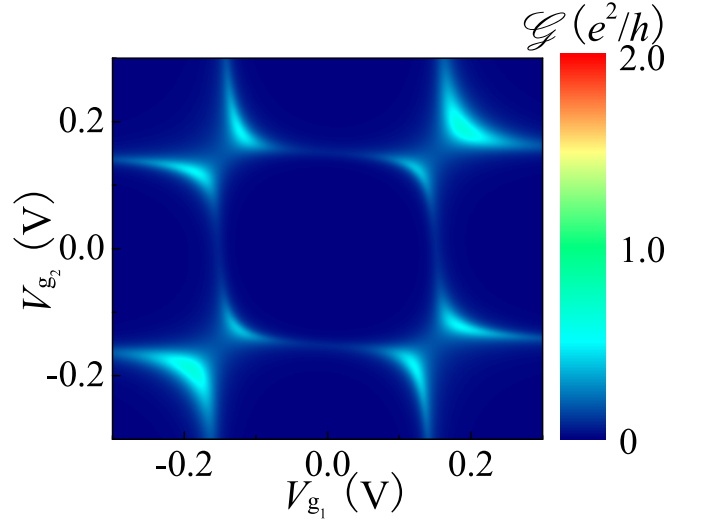,width=1.15 \columnwidth}}
\caption{\label{stab-diagram-DS2}(color) The stability diagram of
serial \DSJ~with
$\epsilon_{1,\sigma}=\epsilon_{2,\sigma}=-0.15~$eV,
$U_{1}=U_{2}=0.3~$eV, $t=0.05~$eV,
$\gamma^{1}_{\textrm{L}}=\gamma^{2}_{\textrm{R}}=0.02~$eV,
$\gamma^{2}_{\textrm{L}}=\gamma^{1}_{\textrm{R}}=0$ ,$
V_{\textrm{bias}}=0.005V$.}
\end{figure}
\subsection{Double site case}
We now investigate the \DSJ\ system with Coulomb interaction on
each dot site. The Hamiltonian is expressed as follows, \beann
H=H_{\textrm{D}}+H_{t}+\sum_{\alpha}(H_{\alpha}+H_{\alpha
\textrm{D}}), \eeann where \beann
&&H^{\phantom{\dagger}}_{\textrm{D}}=\sum_{i,\sigma}\epsilon^{\phantom{\dagger}}_{i,\sigma}d^{\dagger}_{i,\sigma}d^{\phantom{\dagger}}_{i,\sigma}
+\frac{U_{i}}{2}n^{\phantom{\dagger}}_{i,\sigma}n^{\phantom{\dagger}}_{i,\bar{\sigma}},\\
&&H^{\phantom{\dagger}}_{t}=\sum_{i\neq
j,\sigma}\frac{t}{2}(d^{\dagger}_{i,\sigma}d^{\phantom{\dagger}}_{j,\sigma}
+d^{\dagger}_{j,\sigma}d^{\phantom{\dagger}}_{i,\sigma}),\\
&&H^{\phantom{\dagger}}_{\alpha,\sigma}=\sum_{k,\sigma}\epsilon^{(\alpha)}_{k,\sigma}
c^{\dagger}_{\alpha,k,\sigma}c^{\phantom{\dagger}}_{\alpha,k,\sigma},\\
&&H^{\phantom{\dagger}}_{\alpha \textrm{D},
\sigma}=\sum_{k,\sigma}\left(V^{\phantom{\dagger}}_{\alpha,k,\sigma}
c^{\dagger}_{\alpha,k,\sigma}d^{\phantom{\dagger}}_{i,\sigma}
+V^{*\phantom{\dagger}}_{\alpha,k,\sigma}d^{\dagger}_{i,\sigma}
c^{\phantom{\dagger}}_{\alpha,k,\sigma}\right),
\eeann
with $i,j=1,2$ indicate the dot site, $t$ is the
constant for electron hopping between different
sites.
\\
With the help of the EOM, and by means of the truncation
approximation on the double-particle GFs, we obtain the closed
form for the retarded GFs as follows
\begin{subequations}
\begin{align}
\label{EoM-DS-1}
&(\omega-\epsilon^{\phantom{\dagger}}_{i,\sigma}-\Sigma^{\textrm{r}}_{i,\sigma})
G^{(U,t)\textrm{r}}_{i,\sigma;j,\tau}\nonumber\\
&=\delta^{\phantom{\dagger}}_{i,j}\delta^{\phantom{\dagger}}_{\sigma,\tau}
+U_{i}G^{(2)(U,t)\textrm{r}}_{i,\sigma;j,\tau}
+t~G^{(U,t)\textrm{r}}_{i,\sigma;j,\tau},\\
\label{EoM-DS-2}
&(\omega-\epsilon^{\phantom{\dagger}}_{i,\sigma}-U_{i}-\Sigma^{\textrm{r}}_{i,\sigma})
G^{(2)(U,t)\textrm{r}}_{i,\sigma;j,\tau}\nonumber\\
&=\langle
n^{\phantom{\dagger}}_{i,\bar{\sigma}}\rangle\delta^{\phantom{\dagger}}_{i,j}\delta^{\phantom{\dagger}}_{\sigma,\tau}+t~
n^{\phantom{\dagger}}_{i,\bar{\sigma}}G^{(U,t)\textrm{r}}_{i,\sigma;j,\tau},
\end{align}
\end{subequations}
where the DSJ retarded GFs are defined as $
G^{(U,t)\textrm{r}}_{i,j;\sigma,\tau}=\langle\langle
d^{\phantom{\dagger}}_{i,\sigma}|
d^{\dagger}_{j,\tau}\rangle\rangle^{\textrm{r}}$, $
G^{(2)(U,t)\textrm{r}}_{i,j;\sigma,\tau}=\langle\langle
n^{\phantom{\dagger}}_{i,\bar{\sigma}}d^{\phantom{\dagger}}
_{i,\sigma}|d^{\dagger}_{j,\tau}\rangle\rangle^{\textrm{r}} $.
$\bar{i}$ means `NOT $i$', and $\Sigma^{\textrm{r}}_{i, \sigma}$
are the electron self-energy from leads.
\\
From Eqs.~(\ref{EoM-DS-1}), (\ref{EoM-DS-2}) and performing the
same \textit{Ansatz} as in the case of \SSJ, we can obtain the
\DSJ\ lesser GFs with Coulomb-interaction effects as follows, \bea
\label{eq:Gkeldysh-DS-new}
&&G^{(U,t)<}(\omega)=(1+G^{(U,t)\textrm{r}}\Sigma^{\textrm{r}}_{t})
G^{(U)<}\nonumber\\
&&\hspace{0.5cm}\cdot(1+\Sigma^{\textrm{a}}_{t}G^{(U,t)\textrm{a}})
+G^{(U,t)\textrm{r}}\Sigma^{<}_{t}G^{(U,t)\textrm{a}}, \eea with
\beann \Sigma^{\textrm{r}}_{t}=\Sigma^{\textrm{a}}_{t}=
\begin{pmatrix}
0 & t & 0 & 0\\
t & 0 & 0 & 0\\
0 & 0 & 0 & t\\
0 & 0 & t & 0
\end{pmatrix},
\eeann and $\Sigma^{<}_{t}=0$. $G^{(U)<}$ is the DSJ lesser GF
with the same form as Eq.~(\ref{eq:Gkeldysh-SS-new}), but taking
\beann && U=\begin{pmatrix}
U_{1} & 0 & 0 & 0\\
0 & U_{2} & 0 & 0\\
0 & 0 & U_{1} & 0\\
0 & 0 & 0 & U_{2}
\end{pmatrix},\hspace{0.2cm}
\Gamma_{\alpha}=
\begin{pmatrix}
\gamma^{1}_{\alpha} & 0 & 0 & 0\\
0 & \gamma^{2}_{\alpha} & 0 & 0\\
0 & 0 & \gamma^{1}_{\alpha} & 0\\
0 & 0 & 0 & \gamma^{2}_{\alpha}
\end{pmatrix},
\eeann where $\gamma^{i}_{\alpha}$ indicates the line width
function of lead $\alpha$ to site $i$, and $U_{i}$ is the charging
energy at site $i$. $G^{\textrm{r}/\textrm{a}}$ and
$G^{(2)\textrm{r}/\textrm{a}}$ are the GF matrix from
Eqs.~(\ref{EoM-DS-1}) and (\ref{EoM-DS-2}). Here, in order to
distinguish different GFs, we introduce the subscript `$(U,t)$'
for the one with both Coulomb interaction $U$ and inter-site
hopping $t$, while `$(U)$' for the one only with Coulomb
interaction.
\\
For our models, the lesser GFs in Eq.~(\ref{eq:Gkeldysh-SS-new}),
(\ref{GF2-SS}) and (\ref{eq:Gkeldysh-DS-new}), which are obtained
with help of our \emph{Ansatz}, can also be obtained by the EOM
NEGF formula in Eq.~(\ref{new_K-GF}) or in
Ref.~\onlinecite{NiuEOM99jpcm} within the same truncation
approximation.
\section{Transport observables for the double site junction}
The current can be generally written as \cite{Meir92prl} \beann
&&J=\frac{\textrm{i}e}{2\hbar}\int{\frac{d\epsilon}{2\pi}}
\textrm{Tr}\{(\Gamma_{\textrm{L}}-\Gamma_{\textrm{R}})G^{(U,t)<}\nonumber\\
&&\hspace{0.5cm}+[f_{\textrm{L}}(\omega)\Gamma_{\textrm{L}}
-f_{\textrm{R}}(\omega)\Gamma_{\textrm{R}}](G^{(U,t)\textrm{r}}-
G^{(U,t)\textrm{a}})\}, \eeann where the lesser GF is given by
Eq.~(\ref{eq:Gkeldysh-DS-new}). The differential conductance is
defined as \beann \mathcal{G}=\frac{\partial J}{\partial
V_{\textrm{bias}}}, \eeann where the bias voltage is defined as
$V_{\textrm{bias}}=(\mu_{\textrm{R}}-\mu_{\textrm{L}})/e$.
\begin{figure}
\centerline{\psfig{file=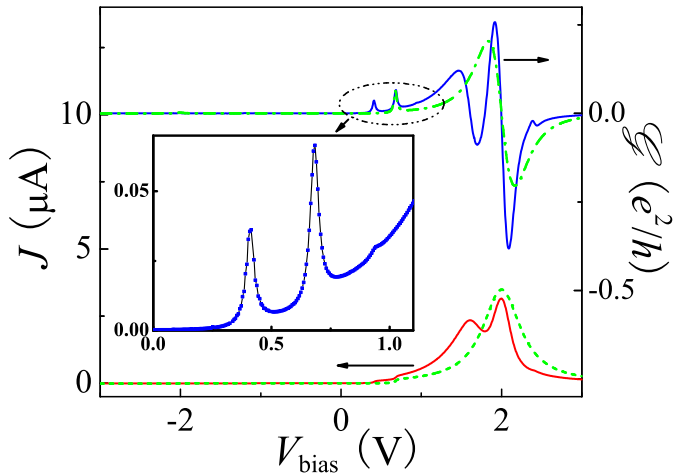,width=1.15\columnwidth}}
\caption{\label{stab-diagram-DS3}(color) Current and conductance
vs.~bias-voltage of a DSJ far from equilibrium with parameters
$\epsilon_{1,\sigma}=0.5~$eV, $\epsilon_{2,\sigma}=-0.5~$eV,
$U_{1}=U_{2}=U=0.2~$eV, $t=0.07~$eV,
$\gamma^{1}_{\textrm{L}}=\gamma^{2}_{\textrm{R}}=0.03~$eV,
$V_{\textrm{g}_{2}}=-V_{\textrm{g}_{1}}=V_{\textrm{bias}}/4$ and
$V_{\textrm{R}}=-V_{\textrm{L}}=V_{\textrm{bias}}/2$. The red
curve represents the current, while the blue the conductance. The
inset is the blow-up for the conductance peak split. The dash and
dot-dash curves are for current and conductance with $U=0$,
respectively.}
\end{figure}
\subsection{Serial configuration}
\begin{figure}[t]
\centerline{\psfig{file=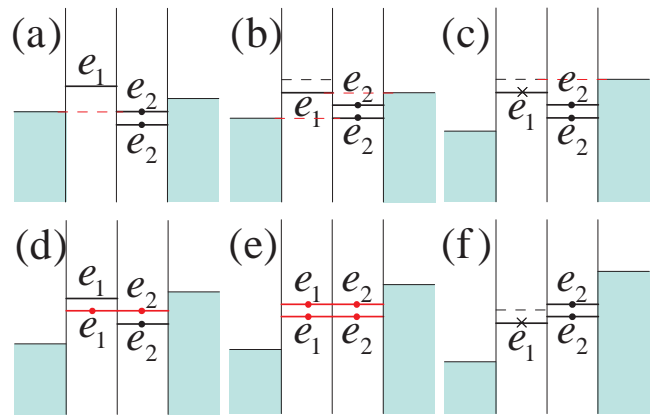,width=1.0\columnwidth}}
\caption{\label{fig-Dynamic}(color) The processes involved in the
transport characteristics in figure \ref{stab-diagram-DS3}.
$e_{1}\equiv\epsilon_{1,\sigma}$,
$e_{2}\equiv\epsilon_{2,\sigma}$,
$e^{\ast}_{1}\equiv\epsilon_{1,\sigma}+U$,
$e^{\ast}_{2}\equiv\epsilon_{2,\sigma}+U$. The red line indicates
electron resonant-tunnelling. a) The first conductance peak. b)
The second conductance peak. c) The pseudo-peak of conductance. d)
The first current maximum, and the red line indicates resonant
tunnelling of electrons. e) The second current maximum for
electron resonant tunnelling. f) The dip of conductance.}
\end{figure}

By taking $\gamma^{2}_{L}=\gamma^{1}_{R}=0$, we
obtain a serial DSJ, which could describe the
kind of molecular quantum dot junctions like the
ones in Ref.~\onlinecite{Elbing05pnas}. First,
at small bias voltages, the conductance with the
two gate voltages $V_{\textrm{g}_{1}}$ and
$V_{\textrm{g}_{2}}$ was calculated, and the
relative stability diagram was obtained as shown
in Fig.~\ref{stab-diagram-DS2}. Because of the
double degeneracy (spin-up and spin-down)
considered for each site and electrons hopping
between the dots, there are eight
resonance-tunnelling regions. This result is
consistent with the master-equation
approach.\cite{DS-review}
\\
Further, we studied the nonequilibrium current for large
bias-voltages (Fig.~\ref{stab-diagram-DS3}). Because
$\epsilon_{1,\sigma}$ and $\epsilon_{2,\sigma}$ are taken as
asymmetric, for the case without Coulomb interaction, the $I$-$V$
curve is asymmetric for $\pm V_{\textrm{bias}}$, and there are one
step and one maximum for the current. The step contributes to one
peak for the conductance. When we introduce the Coulomb
interaction to the system, the one conductance peak is split into
several: two peaks, one pseudo-peak and one dip, while the current
maximum comes to be double split (see
Fig.~\ref{stab-diagram-DS3}). This process can be understood by
the help of Fig.~\ref{fig-Dynamic}. At zero bias-voltage,
$\epsilon_{2,\sigma}$ is occupied and $\epsilon_{1,\sigma}$ is
empty. a) By adding a bias voltage, the Fermi window is opened.
The level $\epsilon_{2,\sigma}+U$ is first resonant with the edge
of the window. It will contribute the first peak for conductance.
b) By opening the window further, the levels $\epsilon_{2,\sigma}$
and $\epsilon_{1,\sigma}$ come into the window resulting in the
second peak. c) When the level $\epsilon_{1,\sigma}+U$ comes in,
only a pseudo-peak appears. This is because there is only a little
possibility for electrons to occupy the level
$\epsilon_{1,\sigma}$ under positive bias voltage. d) Levels
$\epsilon_{2,\sigma}+U$ and $\epsilon_{1,\sigma}$ meet, which
results in electron resonant-tunnelling and leads to the first
maximum of the current. Then a new level $\epsilon_{1,\sigma}+U$
appears over the occupied $\epsilon_{1,\sigma}$ for charge
effects. e) In Fig.~4(e), the meeting of $\epsilon_{2,\sigma}$ and
$\epsilon_{1,\sigma}$ results in electron resonant tunnelling. It
means that $\epsilon_{1,\sigma}$ will be occupied, which leads to
the appearance of a new level $\epsilon_{1,\sigma}+U$. Then
$\epsilon_{2,\sigma}+U$ meets $\epsilon_{1,\sigma}+U$ and another
resonant tunnelling channel is opened for electrons. The two
channels result in the second current maximum. f) Fig.~4(f) shows
that the level $\epsilon_{1,\sigma}+U$ disappears if the level
$\epsilon_{1,\sigma}$ is empty. This means that a dip appears in
the conductance.
\\
It should be noted that the characteristics of
serial DSJ in Fig.~\ref{stab-diagram-DS3} have
showed some reasonable similarities to
experiments of a single-molecule
diode.\cite{Elbing05pnas}
\subsection{Parallel configuration}
\begin{figure}
\centerline{\psfig{file=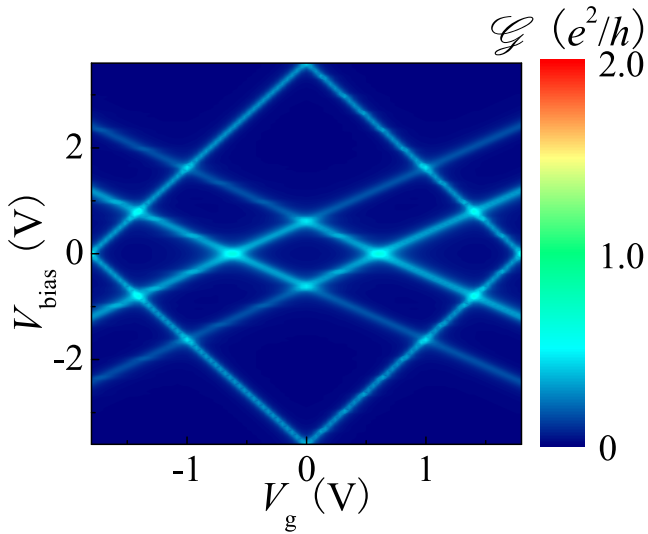, width=1.15\columnwidth}}
\caption{\label{nesting}(color) Nested stability diagram of a
parallel DSJ with parameters $\epsilon_{1,\sigma}=-1.8~$eV,
$\epsilon_{2,\sigma}=-0.3~$eV, $U_{1}=3.6~$eV, $U_{2}=0.6~$eV
$t=0.001~$eV,
$\gamma^{1}_{\textrm{L}}=\gamma^{1}_{\textrm{R}}=0.04~$eV,
$\gamma^{2}_{\textrm{L}}=\gamma^{2}_{\textrm{R}}=0.05~$eV,
$V_{\textrm{g}_{2}}=V_{\textrm{g}_{1}}/2=V_{\textrm{g}}/2$ and
$V_{\textrm{R}}=-V_{\textrm{L}}=V_{\textrm{bias}}/2$.}
\end{figure}

If on the other hand, the two sites are
symmetrically connected to the electrodes,
possibly with a small inter-dot hopping, but
with different charging energies $U_{1}$ and
$U_{2}$, then the stability diagram would show
some nesting characteristics
(Fig.~\ref{nesting}).
\\
The physics of the thin lines in the figure can
be understood by the help of charging effects.
For simplicity, here we would ignore the site
index $i$. In the region of large positive gate
voltage at zero bias voltage,
$\epsilon_{\uparrow}$ and
$\epsilon_{\downarrow}$ are all empty, which
means that the two levels are degenerate.
Therefore adding a bias voltage, first, there
will be two channels ($\epsilon_{\uparrow}$ and
$\epsilon_{\downarrow}$) opened for current
(thick lines). After then, one level
$\epsilon_{\sigma}$ (spin-up or spin-down) is
occupied, while the other obtains a shift for
Coulomb interaction:
$\epsilon_{\bar{\sigma}}\rightarrow
\epsilon_{\bar{\sigma}}+U$. Therefore, when the
bias voltage is further increased to make the
Fermi-window boundary meeting level
$\epsilon_{\bar{\sigma}}+U$, only one channel is
opened for the current, which results in the
thin lines in Fig.~\ref{nesting}. The similar
case appears in the region of large negative
gate voltages.
\section{Conclusions}
In this paper, we introduced a powerful
\textit{Ansatz} for the lesser Green function,
which is consistent with both the Dyson-equation
approach and the equation-of-motion approach. By
using this \textit{Ansatz} together with the
standard equation-of-motion technique for the
retarded and advanced Green functions, we
obtained the NEGF for both the single and the
double site junctions in the Coulomb blockade
regime \emph{at finite voltages} and calculated
the transport observables. The method can be
applied to describe self-consistently transport
through single molecules with strong Coulomb
interaction and arbitrary coupling to the leads.
\\
To test our method, we analyzed the CB stability diagram for a SSJ
and a DSJ. Our results are all consistent with the results of
experiments and the master-equation approach. We showed, that the
improved lesser Green function gives better results for weak
molecule-to-contact couplings, where a comparison with the master
equation approach is possible.
\\
For the serial configuration of a DSJ, such as a donor/acceptor
rectifier, the $I$-$V$ curves maintain diode-like behavior, as it
can be already inferred by coherent transport calculations.
\cite{florian06} Besides, we predict that as a result of the
charging effects, one conductance peak will be split into three
peaks and one dip, and one current maximum into two. For a DSJ
parallel configuration, due to different charging energies on the
two dot sites, the stability diagrams show peculiar nesting
characteristics. In both cases, we present the results of
numerical calculations as well as the simple qualitative picture
of physical processes.
\\
We believe, that the results presented here, beyond their
experimental relevance, might be the transparent base for a
systematic and modular inclusion of a richer physical
phenomenology. Work is currently in progress to include the
electron-vibron interactions to this theory.
\\
\section{Acknowledgments}
We acknowledge fruitful discussions with Rafael Gutierrez and
Florian Pump. This work was funded by the Volkswagen Foundation
under grant No.~I/78~340, by the Deutsche Forschungsgemeinschaft
within the Priority Program SPP 1243, and the trilateral project
CU 44/3-2. Support from the Vielberth Foundation is also
gratefully acknowledged.
\begin{appendix}
\section{Derivation of the EOM lesser Green function}
From the view of perturbation theory, our Hamiltonian can be
generally written as $H=H_{0}+H_{1}$, where $H_{1}$ is the
perturbation term to the solved $H_{0}$. The contour-ordered GF is
defined by means of the Schwinger-Keldysh time contour \bea
\langle\langle
A(\tau_{1});B(\tau_{2})\rangle\rangle^{\textrm{C}}=-\textrm{i}\langle
T_{\textrm{C}}\{A(\tau_{1})B(\tau_{2})\}\rangle, \eea where
$A(\tau_{1})$ and $B(\tau_{2})$ are Heisenberg operators, defined
along the contour C. Taking the time derivative, we obtain the EOM
as, \beann
\textrm{i}\frac{\partial}{\partial\tau_{1}}\langle\langle
A(\tau_{1});B(\tau_{2})\rangle\rangle^{\textrm{C}}&=&\delta^{\textrm{C}}(\tau_{1}-\tau_{2})
\langle[A(\tau_{1}),B(\tau_{2})]_{\pm}\rangle\\ &+&\langle\langle
[A(\tau_{1}),H_{1}];B(\tau_{2})\rangle\rangle^{\textrm{C}}. \eeann
Using the free particle solution
$g^{\textrm{C}}(\tau_{1}-\tau_{2})$, we can rewrite the
time-dependent solution as \beann &&\langle\langle
A(\tau_{1});B(\tau_{2})\rangle\rangle^{\textrm{C}}=g^{\textrm{C}}(\tau_{1}-\tau_{2})
\langle[A(\tau_{1}),B(\tau_{2})]_{\pm}\rangle\\
&&\hspace{0.5cm}+\int{g^{\textrm{C}}(\tau_{1}-\tau')\langle\langle
[A(\tau'),H_{1}];B(\tau_{2})\rangle\rangle^{\textrm{C}}d\tau'}.
\eeann Now applying the Langreth theorem and transforming in the
spectral space, we get \bea \label{new_K-GF} \langle\langle
A|B\rangle\rangle
_{\omega}^{<}&=&g^{<}(\omega)\langle[A,B]_{\pm}\rangle
+g(\omega)^{\textrm{r}}\langle\langle[A,H_{1}],B\rangle\rangle^{<}_{\omega}\nonumber\\
&+&g(\omega)^{<}\langle\langle[A,H_{1}],B\rangle\rangle^{\textrm{a}}_{\omega}.
\eea
\end{appendix}

\end{document}